Title:

# Domain Switching on the Pareto Front: Multi-Objective Deep Kernel Learning in Automated Piezoresponse Force Microscopy

Authors:

Yu Liu[1]*, Utkarsh Pratiush[1], Kamyar Barakati[1], Hiroshi Funakubo[2], Ching-Che Lin[3,4], Jaegyu Kim[4,5], Lane W. Martin[4,6], and Sergei V. Kalinin[1,5]*

[1] Department of Materials Science and Engineering, University of Tennessee, Knoxville, Tennessee, 37996, USA
[2] Department of Material Science and Engineering, Institute of Science Tokyo, Yokohama 226-8502, Japan
[3] Applied Science and Technology Graduate Group, University of California, Berkeley, Berkeley, California 94720, USA
[4] Rice Advanced Materials Institute, Rice University, Houston, Texas, 77251, USA
[5] Department of Materials Science and Engineering, University of California, Berkeley, Berkeley, California 94720, USA
[6] Departments of Materials Science and NanoEngineering, Chemistry, and Physics and Astronomy, Rice University, Houston, Texas, 77251, USA
[7] Physical Sciences Division, Pacific Northwest National Laboratory, Richland, Washington, 99354, USA

* Corresponding author: yliu206@utk.edu, sergei2@utk.edu

## Abstract

Ferroelectric polarization switching underpins the functional performance of a wide range of materials and devices, yet its dependence on complex local microstructural features renders systematic exploration by manual or grid-based spectroscopic measurements impractical. Here, we introduce a multi-objective kernel-learning workflow that infers the microstructural "rules" governing switching behavior directly from high-resolution imaging data. Applied to automated piezoresponse force microscopy (PFM) experiments, our framework efficiently identifies the key relationships between domain-wall configurations and local switching kinetics, revealing how specific wall geometries and defect distributions modulate polarization reversal. Post-experiment analysis projects abstract reward functions—such as switching ease and domain symmetry—onto physically interpretable descriptors including domain configuration and proximity to boundaries. This enables not only high-throughput active learning, but also mechanistic insight into the microstructural control of switching phenomena. While demonstrated for ferroelectric domain switching, our approach provides a powerful, generalizable tool for navigating complex, non-differentiable design spaces, from structure-property correlations in molecular discovery to combinatorial optimization across diverse imaging modalities.

I. Introduction

Materials properties emerge from the collective behavior of features at the micro- and nano-scale. For instance, electrical conductivity in composite systems hinges on the percolation network formed by conductive and insulating phases, while the optical response of nanocomposites is dictated by the size, shape, and distribution of their inclusions.[1, 2] In ferroelectrics, electromechanical coupling arises from the arrangement and dynamics of ferroelectric and ferroelastic domains, structural and topological defects, as well as charged defects within the crystal lattice[3, 4]. Across these examples, it is the underlying microstructure—local defects, grain boundaries, phase interfaces, domain walls—that ultimately controls macroscopic functionality, highlighting the need for quantitative models that connect structural descriptors to performance metrics.[5-8]

In many practical applications, different functional properties are interdependent—sometimes synergistically, often antagonistically. Maximizing one target (e.g., dielectric permittivity) may come at the expense of another (e.g., mechanical stiffness), giving rise to trade-offs that must be carefully managed.[9-11] Capturing these inter-property correlations and mapping out the Pareto front of achievable performance combinations is critical both for materials optimization, where one seeks the best compromise for a given application, and for extracting physical understanding, where the shape of the Pareto frontier reveals fundamental constraints imposed by material chemistry and microstructural physics.

When a large dataset of material properties is readily available, constructing the corresponding Pareto front is straightforward: one simply identifies the outer hull of the observed property vectors.[12-15] In many practical settings, however, while microstructural data may be abundant, each functional measurement (e.g., hysteresis loop acquisition or electron microscopy characterization) is time-consuming and costly.[16-19] In some cases, the functional measurement is even destructive to the structural features, making it intrinsically impossible to map out the local structure-property relation with grid-measurements. Under these constraints, active-learning strategies become essential. Such frameworks selectively propose new experiments, guiding measurements toward microstructures that are most likely to lie on or near the true Pareto frontier, thereby minimizing the number of costly measurements required to map out optimal tradeoffs. In prior work, we introduced deep kernel learning (DKL) as an effective approach for active learning of single-objective material functionalities, demonstrating its utility in both electron[20, 21] and scanning-probe microscopy[22-24] contexts. Here, we extend that methodology to the multi-objective domain. These objectives can be defined as measurable physical properties, such as piezoresponse and switching ease for ferroelectrics and superconducting gap size for superconductors, to directly reflect the goal of experimentation. They can also be defined as physical descriptors extracted from structural maps or experimental parameters, like distance of switching point away from domain boundaries for ferroelectrics and local density of defects for superconductors, to map out the relations between physical properties and these descriptors.

Piezoresponse force microscopy (PFM) presents an almost ideal model system for validating different machine-learning workflows,[25] thanks to the intrinsic quantitative nature of the PFM signal acquired with resonance-tracking modes[26, 27] and the ability to acquire multiple complementary datasets.[28] For example, PFM-based spectroscopies can map hysteresis loops and extract parameters such as nucleation bias, remnant polarization, loop area, and switching nonlinearity.[29, 30] Likewise, nonlinear-spectroscopy measurements yield higher-order coefficients (e.g., second and third-order coefficients reflect the mobility of domain walls) and their derivatives,[31, 32] and when implemented in PFM yield rich information on local electromechanical responses.[33-35]

Here, the quest of discovering domain switching dynamics in ferroelectrics is chosen to challenge our automated active learning workflow. Understanding the dependence of domain-switching behaviors on the underlying domain structure is critical to understand the physics governing the domain switching and to fabricate practical next-generation memory and microelectromechanical systems (MEMS) based on ferroelectric materials. Two ferroelectric materials with distinct microstructural landscapes were chosen: $(PbTiO_3)_{1-x}$-$(KTaO_3)_x$ (PTKTO)[19] and $PbZr_{0.2}Ti_{0.8}O_3$ (PZTO) (111)-oriented epitaxial films.[36-38] PTKTO features pronounced microstructural heterogeneity, including complex and diverse domain wall geometries, which pose a greater challenge for model generalization. Accurate prediction in this setting requires the model to learn subtle correlations between domain boundaries and switching behavior. Conversely, PZTO represents a structurally uniform control system characterized by regularly stripe-like domains. Due to the minimal variation in local structure, PZTO serves as a comparison system as the reward distributions in PZTO are expected to be relatively flat and the exploration trajectory is expected to be randomized.

In this work, we developed a multi-objective Bayesian optimization-based deep kernel learning (MOBO-DKL) framework and applied it to investigate direct polarization switching in ferroelectric materials via fully automated active learning experiments. MOBO-DKL dynamically learns structural representations of local domain environments from PFM maps and aligns them with multiple experimentally measured rewards: local piezoresponse, poled domain size, and poled domain symmetry. These rewards respectively quantify the influence of initial domain configuration, the ease of switching, and the degree of intrinsic disorder. Guided by a joint acquisition strategy, the MOBO-DKL model adaptively selects measurement locations to efficiently map the Pareto front of switching behavior, identifying microstructural features that realize optimal trade-offs between competing objectives. This approach not only accelerates the experimental discovery of structure–property relationships but also provides interpretable insights into the physical mechanisms underpinning polarization reversal, demonstrating a scalable path toward autonomous experimentation in complex functional materials.

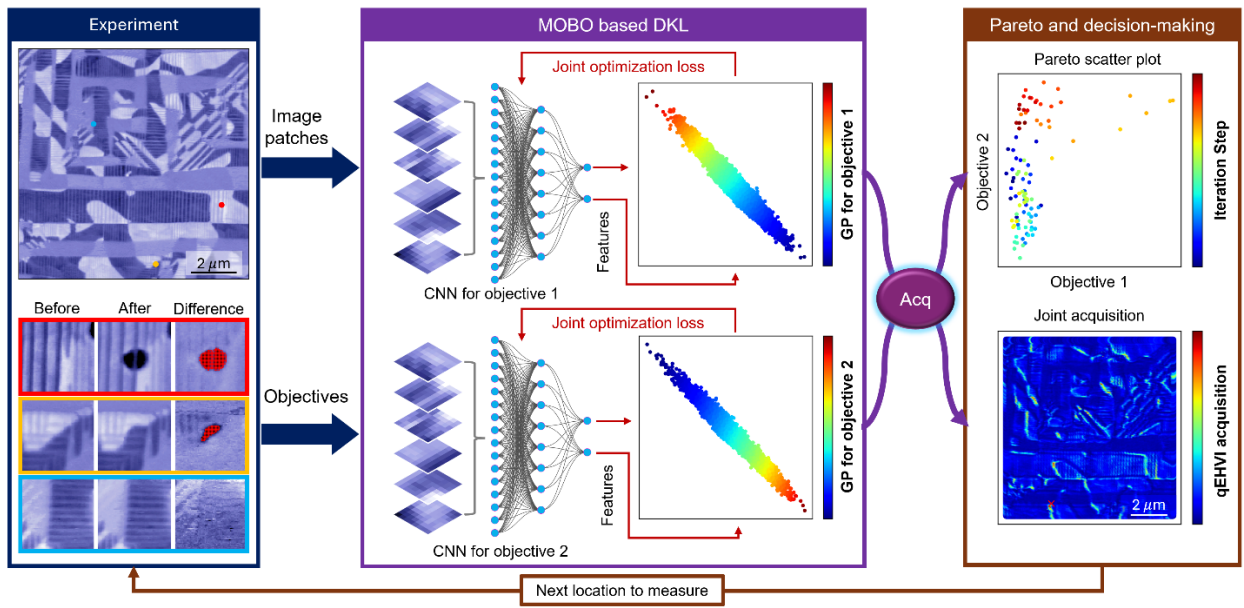

**Figure 1 | Overview of Multi-objective Bayesian optimization (MOBO) based Deep Kernel Learning (DKL).** A MOBO-DKL active learning workflow consists of three parts. (1) Experimental stage: generating structural patches based on global map and extracting objectives from local measurements. (2) Learning stage: CNN-GP joint training on measured patches and their corresponding objectives. (3) Pareto and Decision-making stage.

## II. Results and discussions

A MOBO-DKL active learning workflow consists of three parts (Fig. 1). In the experiment stage, a global map provides cropped image patches as the structural input of MOBO-DKL. Single or multiple measurements at the center of cropped image patches provide physical properties as objectives (rewards) input of MOBO-DKL. At the active learning stage, the cropped image patches will be fed into convolutional neural networks (CNN) to extract key features, which will be fitted by multiple Gaussian processes (GP) with the rewards extracted from physical property measurements centered at corresponding patches. The CNNs and GPs are optimized jointly to make sure the CNNs extract key features that align with the measured rewards. At the decision-making stage, a joint acquisition function is computed with the predictions and uncertainties of all the GPs considered, from which the next location to measure is determined. From the Pareto relation between each reward, the physics behind them can be discovered.

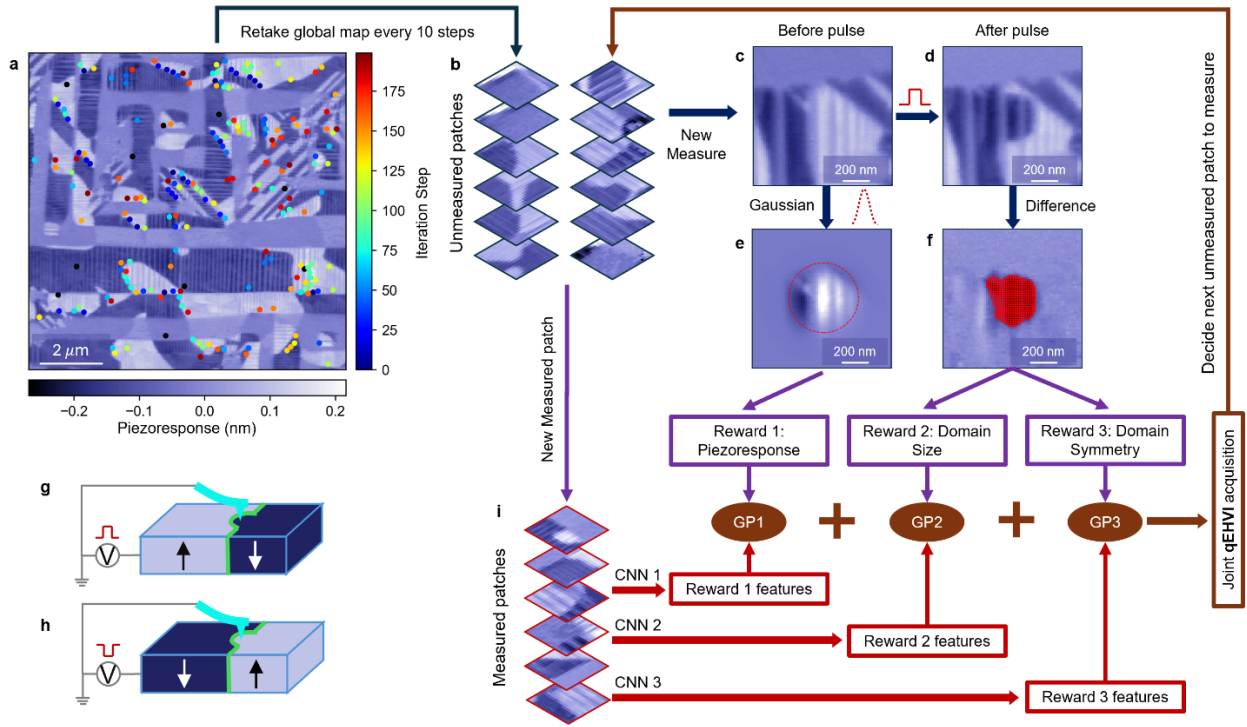

**Figure 2 | Setup and workflow of automated experiment. a,** Overview map of a MOBO-DKL active learning experiment. Here, 10 seeding steps (black dots) plus 200 training steps (rainbow-colored dots) are overlayed on the global piezoresponse map, showing the trajectory of the learning process. **b,** the global map is cropped into unmeasured and measured patches. **c,** a 1 $\mu$m PFM scan records the domain structure before domain switching. **d,** an after-pulse scan after a $\pm 30$ V pulse

is applied for 10 seconds at the center of the scan. **e,** reward of piezoresponse is computed by averaging the before-pulse map in **c** with a Gaussian mask applied at the center. The red-dashed circle represents the width of the Gaussian mask. **f,** the difference between the after-pulse and before-pulse maps shows the switched domain. **g-h,** schematic drawing of switching domain by applying pulse polarity determined by the phase of the before-pulse scan. **i,** the MOBO-DKL model is trained between all the measured patches and their rewards, where the CNNs and GPs are optimized jointly to extract features that align with the measured rewards. A joint q-expected hypervolume improvement (qEHVI) acquisition function is computed to close the active learning iteration loop by determining the next location to measure.

In the MOBO-DKL study of domain poling dynamics in PTKTO, we used the piezoresponse map as the global structural map, which is obtained through PFM scans completed in the dual amplitude resonance tracking (DART) mode (Fig. 2a). The classification of domains according to their piezoresponse is shown in Fig. S1. The (256 x 256) global map is cropped into (16 x 16) patches to highlight local structural features, which are further divided into unmeasured patches and measured patches (Fig. 2b). In the beginning of the experiment, 10 seeding measurements were taken to initialize the model. Each active learning iteration cycle starts with selecting a new location to measure from unmeasured patches. Then a 1 $\mu$m (64 x 64) PFM scan will be taken to record the domain structure before poling the domains at the new location to measure (Fig. 2c). After applying a $\pm 30$ V voltage pulse for 10 seconds, an after-pulse map with identical settings is taken to image the effect of the voltage pulse (Fig. 2d).

Here, we have defined three rewards: 1) the averaged piezoresponse extracted from before-pulse map (Fig. 2e), indicating the pre-existing neighboring domain structure; 2) the poled-domain size measured from the difference map (Fig. 2f), representing how easy it is to switch the domain; and 3) the poled-domain symmetry, which is computed as how far the poled domain has deviated from a perfect cycle and contains information about underlying disorder and pre-existing domain structures. Those rewards are defined as follows:

$$\text{Reward 1} = \overline{P(\vec{r}) \cdot \exp(-\vec{r}/w)},$$

$$\text{Reward 2} = \sqrt{\frac{N_{switched}}{\pi}}, \text{ and}$$

$$\text{Reward 3} = \frac{\overline{(r_{i0})}}{\sigma(r_{i0})}$$

where $P(\vec{r})$ is the piezoresponse at position $\vec{r}$, $w$ is the width of Gaussian mask (indicated as red dashed circle, Fig. 2e), $N_{switched}$ is the area of poled domain (Fig. 2f), $r_{i0}$ is the distance from the $i^{th}$ poled pixel to the center of mass of all poled pixels, and $\sigma(r_{i0})$ is the standard deviation of $r_{i0}$.

After the measurement, the MOBO-DKL model is trained between all the measured patches and their rewards (Fig. 2i), where the CNNs and GPs are optimized jointly to extract features that align with the measured rewards. A joint qEHVI acquisition function is computed based on the predictions and uncertainties of all GPs, which closes the active learning iteration loop by determining the next location to measure for the next cycle. The detailed model construction is included in SI-section 1 and detailed model training process is provided in Methods. We have also provided recorded videos showing the fully automated operation in the attached media file 1.

During the experiment, the global structural map was reacquired every 10 active learning steps, updating only the unmeasured patches. This approach compensates for drift and measurement-induced changes in the domain structure over time. The entire MOBO-DKL experiment was completed in approximately 7 hours for 210 total steps. In contrast, conducting a full-grid measurement to generate distributions of properties with similar resolution as the global map in Fig. 2a would require an estimated three months. Moreover, such an exhaustive approach would still fail to accurately capture the structure–property relationships, as the domain-switching measurements are inherently destructive and can alter the surrounding domain configurations.

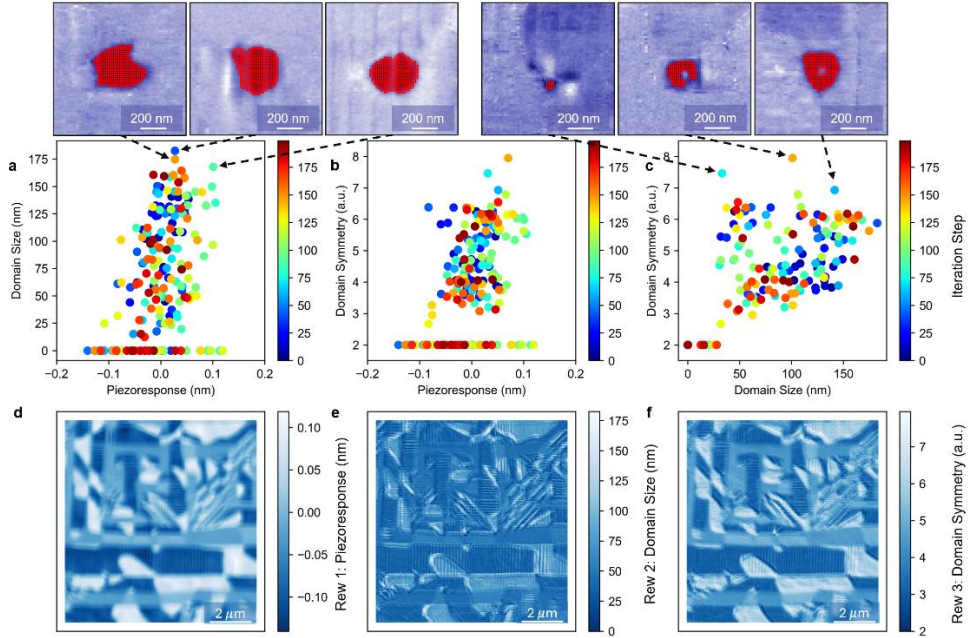

**Figure 3 | Pareto plot and predicted distribution of rewards. a-c,** Pareto scatter plots for **a** poled domain size vs. piezoresponse, **b** poled domain symmetry vs. piezoresponse, and **c** poled domain size vs. poled domain symmetry. Here the results are obtained through the active learning experiment described in Fig. 2. The insets above show the characteristic difference maps after poling for three largest poled domain sizes and three highest poled domain symmetry. **d-f,** predicted distribution of **d** averaged piezoresponse, **e** poled domain size, and **f** poled domain symmetry in the original 10 $\mu$m global map.

At the end of the automated experiment, the results are Pareto scatterplots showing the interdependence between different rewards, and predicted distributions of the rewards based on the global structural map. Here, a Pareto point is a point in the reward (hyperparameter) space such that no solution can be improved in one objective without incurring a cost in another. The trained MOBO-DKL model can be used to extract key structural features from all cropped patches and then predict the distribution of associated rewards across the global map. The trajectory of active learning shown as the full experiment trail is plotted (Fig. 2a), indicating the preference of the model with tradeoffs between all rewards considered. The post-acquisition analysis of the automated experiment can be performed similarly to standard DKL and GP workflows as described in Refs. 39-41

The distribution of poled-domain size, which peaks near zero value (Fig. 3a), indicates that domains are most easily switched near the boundaries between the positive and negative out-of-plane polarization states (blue and red domains, Fig. S1), where Reward 1 of average piezoresponse is close to zero. Moreover, positive domains (with positive piezoresponse) are generally easier to switch than those with negative piezoresponse, as evidenced by a skew in the domain-size distribution toward positive piezoresponse values. These trends are consistent with the observed exploration trajectory, which predominantly samples locations near the positive/negative domain boundaries, particularly on the positive-domain side (Fig. 2a). These conclusions are further supported by post-experiment segmentation analysis (Figs. S1 and S2).

Figure 3c reveals a generally positive correlation between poled-domain size and symmetry, as indicated by the Pareto scatter points distributed along the upper-right quadrant. Notable outliers exist, however, including small but highly symmetric domains, suggesting that while larger domains tend to have higher symmetry, this relationship can be disrupted – likely due to disorder and constraints imposed by nearby domain boundaries. Here, the reward of symmetry is only used as a rough estimation of disorder and anisotropy of pre-existing domains. A more advanced symmetry reward can be constructed, for example, as the orientation angle of the primary axis of the poled domain, which can be a powerful tool to show the directionality of domain switching and their dependence on the domain structures and crystal lattices.

Finally, the distribution of poled-domain size predicted by the trained MOBO-DKL model (Fig. 3e) shows that domain structures can be ranked in terms of switching ease as follows: domain boundaries > positive domains > negative domains > in-plane domains.

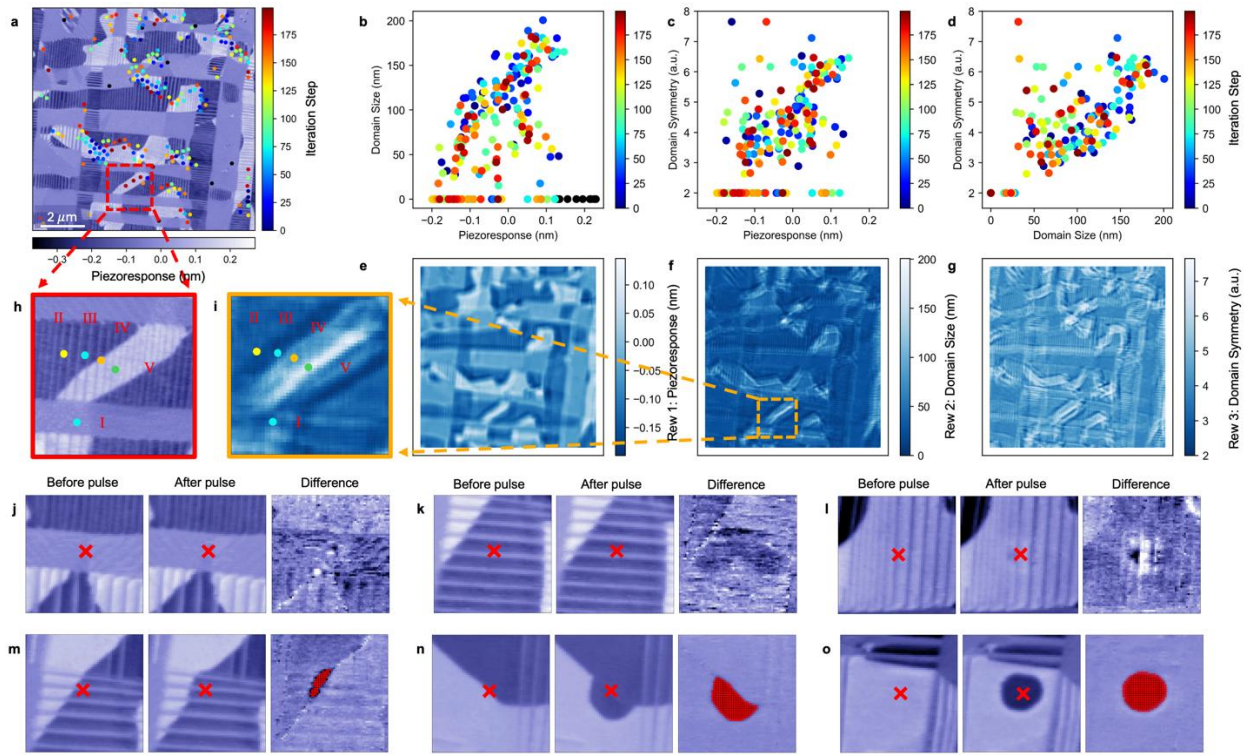

**Figure 4 | Explainable interpretation of MOBO DKL results. a,** Overview map of a MOBO-DKL active-learning experiment. There are 10 seeding steps and 200 active learning steps in this experiment, and the voltage pulse is kept at –30 V for 10 seconds for all measurements. **b-d,** Pareto scatterplots for **b** poled-domain size vs. piezoresponse, **c** poled-domain symmetry vs. piezoresponse, and **d** poled-domain size vs. poled-domain symmetry. **e-f,** predicted distribution of **e** averaged piezoresponse, **f** poled-domain size, and **g** poled-domain symmetry in the original 10 $\mu$m global map. **h-i,** enlarged maps of **h** piezoresponse and **i** predicted poled-domain size. The locations I-V show characteristic domain structures with red crosses marking the location of the applied pulses. **j-o,** location I shows an in-plane domain, which is difficult to switch with out-of-

plane electric field as exampled in **j**. Location II shows the interiors of out-of-plane domains that are far from domain boundaries, which is also difficult to switch as exampled in **k** and **l**. Locations III-V show a pathway across a domain boundary between two out-of-plane domains. The poling behavior changes from shifting the domain boundary that creates a small and asymmetric poled domain, to creating a large and symmetric poled domain as indicated in **m-o**. All these behaviors are captured in the MOBO-DKL predicted poled domain size map in **i**.

To validate the predictions of the MOBO-DKL model, we conducted an additional active learning experiment in which the voltage pulse was fixed at –30 V for 10 seconds. This setup targets the selective switching of positive domains. While some poled-domain sizes with negative piezoresponse were still observed (Fig. 4b), these primarily result from shifting the domain boundaries toward the positive domains rather than switching the negative domains.

Figures 4h–i present a comparison between the structural map and the predicted poled-domain-size map for the same enlarged region. The small predicted poled-domain size at location I (Fig. 4i) results from the measurement shown in Fig. 4j that the in-plane domains are largely unswitchable. Location II lies within the interior of out-of-plane domains, far from domain boundaries, where domain switching requires a higher voltage and is thus limited under the fixed –30 V condition (Figs. 4k-l).

Along the trajectory from Location III to Location V, both the predicted poled-domain size and symmetry increase. This trend reflects the evolution of poling behavior from pushing domain boundaries toward positive domains to inducing large, symmetric negative domains (Figs. 4m–o). Notably, this progression is well captured by the model's predictions in Fig. 4i.

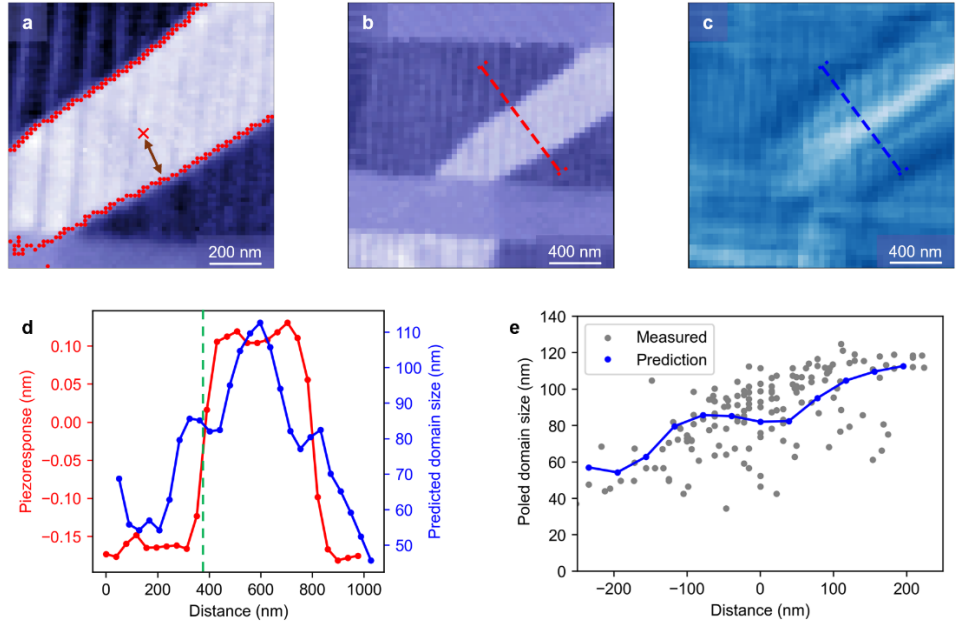

**Figure 5 | Quantitative interpretation of MOBO-DKL predictions. a,** The minimum distance between the location of voltage pulse and its closest domain boundaries (the double-headed arrow) can be computed for each before-pulse piezoresponse map. The positions of all the domain boundaries are located by applying a Canny filter. All the nonzero measured poled domain sizes are scattered with their corresponding minimum distance in **e,** yielding a statistical dependence of poled domain size on the distance to the closest domain boundary. **b-c,** linecuts across an positive domain for **b** piezoresponse map and **c** predicted poled domain size. The linecuts are taken along the red and blue dashed lines, with the segment at the ends representing the width of average perpendicular to the linecuts. **d,** comparison between the two linecuts taken in **b** and **c**. The green dashed line indicates the location of a domain boundary inferred from the inflection point in the piezoresponse linecut. **e,** the linecut of predicted poled domain sizes can be plotted on top of the measured poled domain sizes for a direct comparison after shifting the green dashed line in **d** to zero distance.

To validate the reliability of the MOBO-DKL predictions, we extracted the measured poled-domain size as a function of the minimum distance between the voltage-pulse locations and their nearest domain boundaries (Fig. 5a). Both the measured domain sizes and distances were converted into physical units for comparison. The relationship between the predicted poled-domain size and the distance from the nearest domain boundary was obtained by taking a linecut perpendicular to the boundary. The domain boundary was identified as the inflection point along the linecut in the piezoresponse map (Figs. 5b and 5d). A direct comparison of predicted and measured poled-domain sizes (Fig. 5e) demonstrates good agreement in both magnitude and spatial dependence. This agreement confirms that the MOBO-DKL model can reliably predict complex material responses based solely on structural information extracted from global maps. In addition, the similar distribution of poled-domain size vs. averaged piezoresponse (Fig. 4b) and vs. distance (Fig. 5e) indicates that our proposed reward of averaged piezoresponse is a good measure or physical descriptor of distance from the nearest domain boundary. The introduction of such physical descriptor as a reward into MOBO-DKL experiment not only diversifies the exploration trajectory, but also directly maps out the dependence of switching behavior on the local domain structural variations.

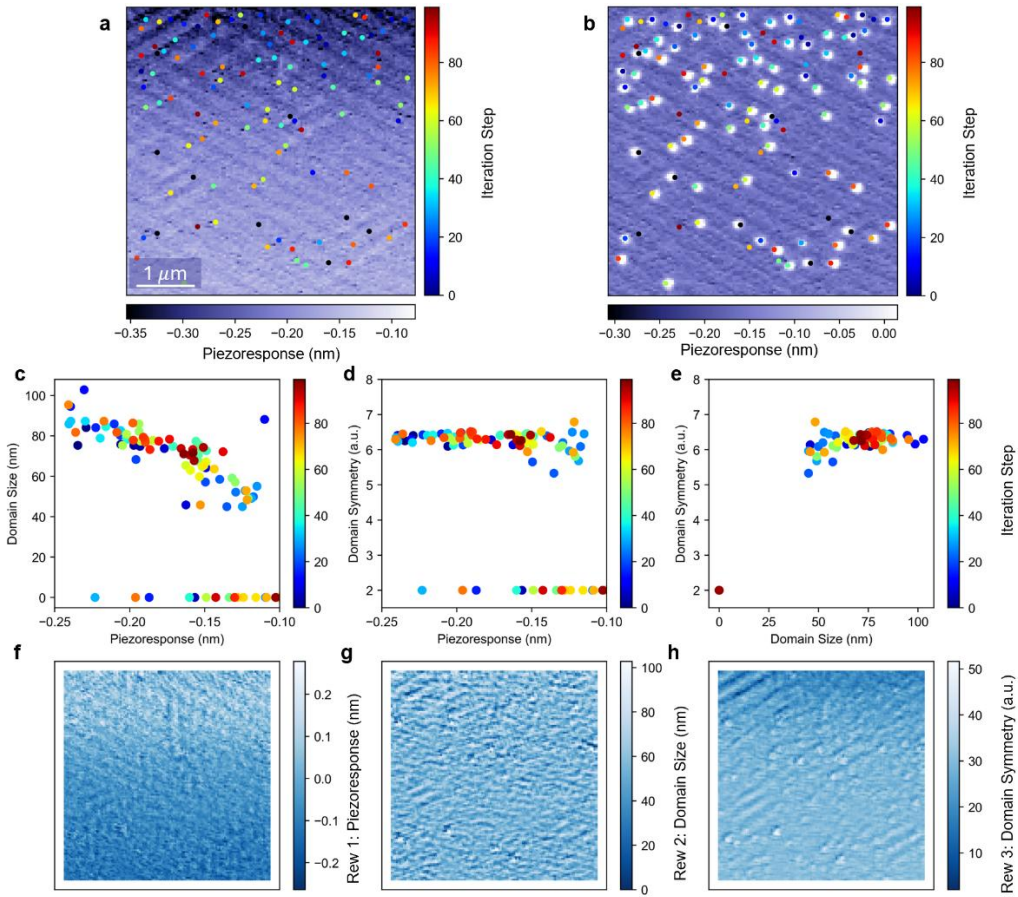

**Figure 6 | MOBO DKL of PZTO sample. a,** The global piezoresponse map before the experiment. Two scans with +8 V and –8V constant voltage applied to the sample are performed to erase the natural domains and form a uniformly poled domain. There are 10 seeding and 100

learning steps, with the switching voltage pulse fixed at +8 V for 10 seconds. **b,** the global map after the active learning experiment, showing the trajectory of all the steps. The white dots indicate the switched domains. **c-e,** Pareto scatter plots for **c** poled domain size vs. piezoresponse, **d** poled domain symmetry vs. piezoresponse, and **e** poled domain size vs. poled domain symmetry. **f-h,** predicted distribution of **f** averaged piezoresponse, **g** poled domain size, and **h** poled domain symmetry in the original 5 $\mu$m global map.

We further explored the behavior of highly uniform (111)-oriented PZTO films to verify the robustness of the MOBO-DKL algorithm. After scanning with +8 V and –8 V constant sample voltage in sequence, the as-grown domain structures are erased and, in their place, forms stripe-like domains uniformly distributed within the 5 $\mu$m map. The resulting MOBO-DKL trajectory shows uniform distribution across the map (Fig. 6a), which suggests the robustness of algorithm and is an important measure of performance. Such a null result shows that when the sample is highly uniform, or the measured properties have little dependence on the structures, the algorithm will behave like random exploration as expected instead of falsely clustering.

Although both the bright and dark stripe domains (Fig. 6a-b) give the same downward out-of-plane domain, their piezoresponse and switching behaviors are different as revealed by the linear dependence of the poled-domain size on the underlying piezoresponse amplitude (Fig. 6c). Dark-stripe domains are easier to switch and give larger piezoresponse, implying that they less constraints on domain flipping. Such observation can be explained by different switching energy and pathways correlated with different in-plane configurations in these two domains.[37, 38] On the other hand, the uniformly distributed poled-domain symmetry across different piezoresponse (Fig. 6d) and different poled-domain size (Fig. 6e) shows that the switching is highly isotropic along all directions.

### III.    Conclusion

In this work, we developed a MOBO-DKL workflow and applied it to uncover the physics governing domain switching in PTKTO and PZT samples through fully automated experiments. Our results demonstrate that MOBO-DKL–driven active learning can effectively capture the structure–property relationships across multiple reward metrics, reveal their interdependence on the Pareto front, and organize complex experimental outcomes into interpretable and actionable insights. The model's predictions—based solely on structural maps—not only reproduce the qualitative evolution of domain switching across varying domain structures, but also yield reliable quantitative trends, particularly with respect to proximity to domain boundaries. The success of the DKL and especially MOBO DKL for automated experiments depends on the correlation between observed structural images and the target functionality. Discovery of such correlation is the goal of the DKL experiments. We also note that equivalent multi-objective workflows can in principle be constructed using simpler dimensionality-reduction techniques. For example, one could apply principal component analysis or a standard variational autoencoder to derive reduced coordinates and then fit a multi-output Gaussian process on those features. Although a systematic comparison of these alternative architectures is beyond the scope of this paper, we release our datasets openly to enable the community to explore and benchmark diverse multi-objective modeling approaches.

Finally, the integration of multiple reward functions emulates the human decision-making process of navigating trade-offs between competing objectives. This approach diversifies the exploration space and prioritizes structures that exhibit nontrivial behavior across all reward dimensions, enhancing both discovery and interpretability. Ultimately, MOBO-DKL provides a powerful and generalizable framework for autonomous experimentation and scientific insight in complex, high-dimensional materials systems. Such framework can be expanded from real microstructures to abstract structures like molecular spaces.

## IV.   Methods

**MOBO-DKL implementation and automated instrument control**

For each objective, we trained a deep kernel Gaussian Process model using a shallow convolutional neural network as the feature extractor. The network, implemented in PyTorch[42], consists of two convolutional layers with ReLU activations and max-pooling, followed by a dynamically initialized fully connected layer that maps each image patch to a low-dimensional latent space. This embedding is passed to a Gaussian Process with a radial basis function (RBF) kernel and constant mean, implemented using GPyTorch. GPyTorch inherits from PyTorch, enabling easy integration with neural networks and modular training of probabilistic models. The GP is trained via stochastic variational inference using a Cholesky variational distribution and inducing points selected from the training data. The loss function is the negative Evidence Lower Bound (ELBO), optimized using Adam (learning rate 0.01) for 50–100 epochs per step. All training is performed in double precision (float64) on GPU (cuda) when available. For query selection, we use the q-Expected Hypervolume Improvement (qEHVI) acquisition function provided by BoTorch[43], a Bayesian optimization library built on top of PyTorch and GPyTorch. BoTorch enables seamless integration of acquisition functions with probabilistic models, supporting discrete optimization over candidate patches to guide the selection of the next query point from the simulated/real microscope.

The SPM control is achieved by our home built open-source Python interface library, AESPM[44]. This library not only enables real-time operating the SPM system local or remotely with code the same way as human operators but also has access to the intermediate data like trace and retrace scan lines in all the channels which are essential for fast optimization presented in this work. We have included a recorded video showing the fully-automated MOBO-DKL experimentation.

AESPM is an open-source SPM-Python interface library. It can be found in the following link with detailed examples and tutorial notebooks: https://github.com/RichardLiuCoding/aespm

To help readers understand and reproduce the results in this work, we have provided an open-source Python notebook of applying MOBO-DKL in a simulated automated experiment: https://github.com/RichardLiuCoding/Publications/blob/main/MOBO_DKL_RL_dev_v2(qEHVI).ipynb

**PFM scan and domain writing**

All the PFM scan and domain writing experiments were conducted on a Jupiter AFM system (Oxford Instrument Asylum Research) with a high-voltage sample stage and head. Fresh Multi75G probes from BudgetSensor were used.

**Chemical vapor deposition growth of PTKTO**

As a model system, epitaxial $PbTiO_3$ ($PTO$) thin films were deposited on (100) $KTaO_3$ single-crystal substrates by pulsed metal–organic chemical vapor deposition. Precursors of $Pb(C_{11}H_{19}O_2)_2$, $Ti(O-i-C_3H_7)_4$ were pulsed in alternation and $O_2$ was continuously introduced, with film thickness tuned simply by varying deposition time. During growth the substrate was held at $600\ °C$ and 670 Pa; afterward it was cooled to room temperature at $10\ °C\ min^{-1}$. The coherent tensile misfit strain (~ 0.5% in 50 nm films, relaxing to ≈ 0% by $1\ \mu m$ thickness) elevates the Curie temperature by ~ $100\ °C$ and stabilizes in-plane polarization.

**Pulsed-laser deposition of thin-film heterostructures**

Heterostructures consisting of 150 nm PbZr$_{0.2}$Ti$_{0.8}$O$_3$ (PZT) and 30 nm La$_{0.67}$Sr$_{0.33}$MnO$_3$ (LSMO) were grown on SrTiO$_3$ (STO) substrates with (001), (110), and (111) orientations (MTI Corp.) using pulsed-laser deposition (PLD). A KrF excimer laser ($\lambda$ = 248 nm, LPX 300, Coherent) was employed to ablate ceramic targets with nominal compositions of Pb$_{1.2}$Zr$_{0.2}$Ti$_{0.8}$O$_3$ and La$_{0.67}$Sr$_{0.33}$MnO$_3$ (Praxair Inc.). Both the PZT and LSMO layers were deposited under identical conditions: a target-to-substrate distance of 60 mm, a substrate heater temperature of 635 °C, an oxygen partial pressure of 200 mTorr, a laser fluence of 1.5 J cm$^{-2}$, and a laser repetition rate of 4 Hz. After deposition, the samples were cooled to room temperature at a rate of 10 °C min$^{-1}$ in flowing oxygen at a pressure of ~700 Torr.

**Acknowledgements**

The development and testing of MOBO-DKL workflow (YL, UP, KB and SVK) was supported by the Center for 3D Ferroelectric Microelectronics (3DFeM), an Energy Frontier Research Center funded by the U.S. Department of Energy (DOE), Office of Science, Basic Energy Sciences under Award Number DE-SC0021118. H. F. was partially supported by the Japan Science and Technology Agency (JST) as part of Adopting Sustainable Partnerships for Innovative Research Ecosystem (ASPIRE)(JPMJAP2312), MEXT Initiative to Establish Next-generation Novel Integrated Circuits Centers (X-NICS) (JPJ011438), and MEXT Program: Data Creation and Utilization Type, Material Research and Development Project (JPMXP1122683430). C.-C.L. acknowledges that this research was sponsored by the Army Research Laboratory and was accomplished under Cooperative Agreement Number W911NF-24-2-0100. J.K. acknowledges Army Research Office under the ETHOS MURI via cooperative agreement W911NF-21-2-0162. The views and conclusions contained in this document are those of the authors and should not be interpreted as representing the official policies, either expressed or implied, of the Army Research Laboratory or the U.S. Government. The U.S. Government is authorized to reproduce and distribute reprints for Government purposes notwithstanding any copyright notation herein. L.W.M. acknowledges the National Science Foundation under Grant DMR-2329111.

# Supplementary Information

## 1. Introduction to the MOBO-DKL algorithm

### a. Gaussian process

A Gaussian Process (GP)[45] is a non-parametric probabilistic model that defines a distribution over functions.

It is specified by a mean function $m(x)$ and a covariance function $k(x, x')$:

$$f(x) \sim \mathcal{GP}(m(x), k(x, x'))$$

Typically, $m(x) = 0$ and the covariance is modelled using an RBF kernel:

$$k(x, x') = \sigma^2 \exp\left(-\frac{|x - x'|^2}{2l^2}\right)$$

Given training data ($\mathcal{D} = \{X, y\}$), the predictive posterior for a new input $x*$ is:

$$f(x*) \mid D \sim N(\mu*, \sigma*^2)$$

With,

$$\mu_* = k_*^\top (K + \sigma_n^2 I)^{-1} y, \quad \sigma_*^2 = k(x_*, x_*) - k_*^\top (K + \sigma_n^2 I)^{-1} k_*$$

### b. Deep kernel Learning (DKL)

DKL combines a neural network for feature extraction with a Gaussian Process as a final probabilistic layer. The idea is to learn a mapping $\phi(x; \theta)$ such that[46]:

$$f(x) \sim \mathcal{GP}\left(0, k(\phi(x; \theta), \phi(x'; \theta))\right)$$

This allows the kernel to operate in a learned feature space instead of the raw input space. Training typically involves maximizing the marginal likelihood with respect to both GP hyperparameters and neural network weights.

$$\log p(y \mid X, \theta) = -\frac{1}{2} y^\top K_\phi^{-1} y - \frac{1}{2} \log|K_\phi| - \frac{n}{2} \log(2\pi)$$

### c. Multi-objective Deep Kernel Learning

**MO-DKL** extends DKL to handle multiple, often conflicting, objectives in the context of **multi-objective optimization (MOO)**.[47] Given objectives $(f_1(x), \dots, f_m(x))$, Here, each objective $f_i$ has its own separate neural network $\phi_i(x; \theta_i)$ followed by its own GP:

$$f_i(x) \sim \mathcal{GP}\left(0, k_i(\phi_i(x), \phi_i(x'))\right)$$

The model is trained on Variational Evidence Lower Bound (ELBO) Loss

$$\mathcal{L} = -E_{q(f)}[\log p(\mathbf{y}|f)] + \text{KL}[q(f)|p(f)]$$

where $\mathcal{L}$ is the total loss to be minimized. The variable $f$ denotes the latent function modeled by the GP, and $\mathbf{y}$ represents the observed outputs. The term $q(f)$ is the variational distribution that approximates the true posterior $p(f/y)$, while $p(f)$ is the GP prior over functions. The first term, $E_{q(f)}[\log p(\mathbf{y}|f)]$ is the expected log-likelihood under the variational distribution and encourages the model to explain the data well. The second term, $\text{KL}[q(f)|p(f)]$, is the Kullback-Leibler (KL) divergence, which penalizes divergence from the prior and serves as a regularizer. Together, these terms balance the trade-off between fitting the data and maintaining a function distribution close to the prior, forming the basis of variational inference in approximate Gaussian Processes.

### d. Pareto optimality and acquisition function

In Active learning we are interested to select to most useful next point and the "usefulness" of the next point can be defined in terms of pareto optimality and acquisition function.

In multi-objective optimization, we are interested in optimizing a vector-valued function $f(x) = [f_1(x), f_2(x), \ldots, f_m(x)]$, where the objectives are often conflicting. A solution $x*$ is said to be **Pareto optimal** if no other solution exists that improves at least one objective without worsening another. Formally, a point x **dominates** another point x' if:

$$\forall i \in \{1, \ldots, m\}, \; f_i(x) \leq f_i(x') \quad \text{and} \quad \exists j, \; f_j(x) < f_j(x')$$

The set of all non-dominated solutions forms the **Pareto front**: a frontier that represents the trade-offs between different objectives.

Expected Hypervolume Improvement (EHVI)[48, 49] is a Bayesian optimization acquisition function designed for multi-objective problems. It quantifies the expected increase in hypervolume that would result from evaluating a new candidate point x. The hypervolume is defined as the volume of objective space dominated by the current Pareto front with respect to a user-defined reference point.

Given a current Pareto set ($\mathcal{P}$), the hypervolume improvement from a new point $f(x)$ is:

$$\Delta HV(x) = HV(\mathcal{P} \cup \{f(x)\}) - HV(\mathcal{P})$$

Since $f(x)$ is not known in advance, we take the expectation over the posterior distribution of the surrogate model (e.g., Gaussian Process), leading to:

$$\text{EHVI}(x) = E_{f(x)}[\Delta HV(x)]$$

EHVI selects the point x with the highest expected gain in hypervolume, favoring evaluations that most efficiently expand the known Pareto front. It is especially effective when combined with probabilistic models like Gaussian Processes, where the uncertainty in predictions can be meaningfully exploited.

2. Manual analysis of the automated experiment results

Figure S1 and S2 show the post-experiment analysis of the MOBO-DKL results in Figure 3.

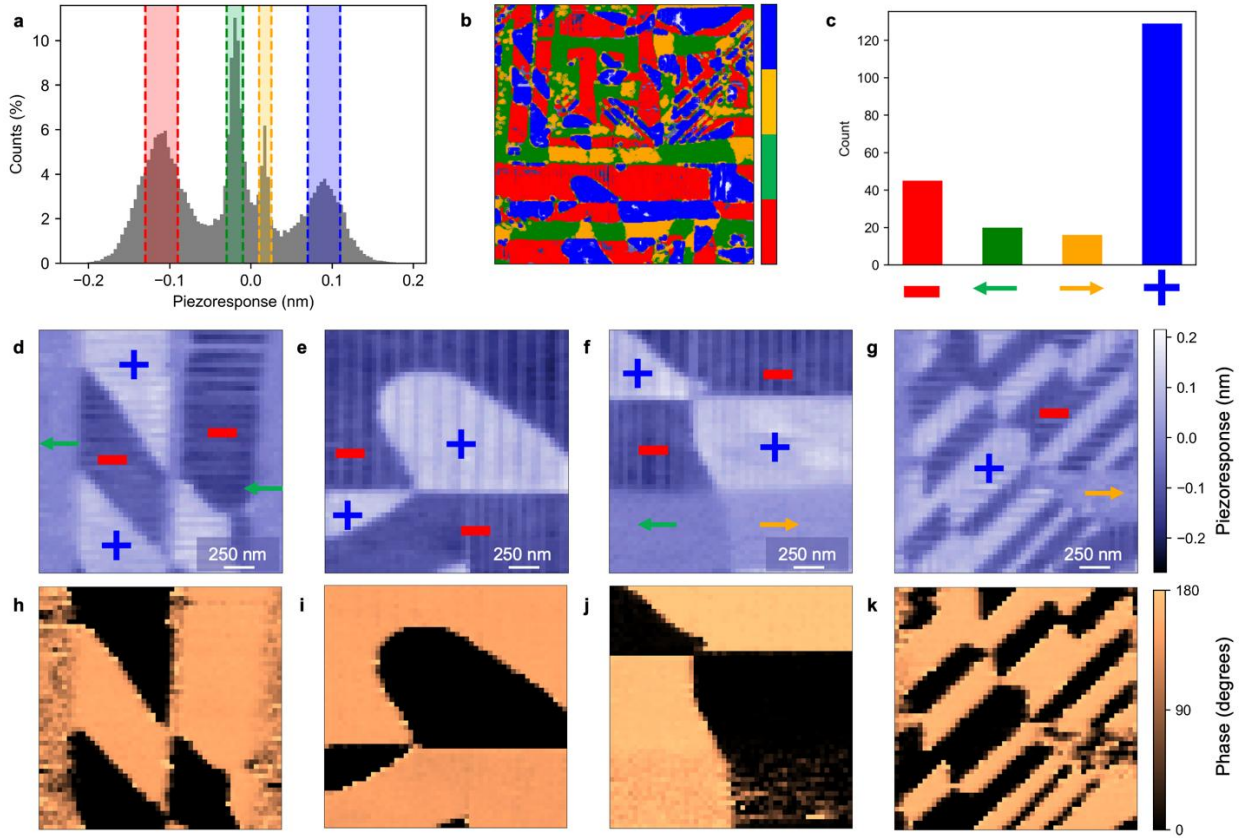

**Figure S1 | Manual segment of the domains. a,** histogram of piezoresponse from the global map in Fig. 2a. **b,** the global map can be segmented into four classes, each represented by a color corresponding to a peak in **a. c,** all the measured patches can be classified by the piezoresponse measured at the center of before-pulse map into positive out-of-plane (blue), negative out-of-plane (red), and in-plane domains (green and orange). **d-g,** four cropped piezoresponse map show characteristic domain boundaries between the four types of domains, with colored marks showing the identity of each domain. **h-k,** phase maps correspond to piezoresponse maps in **d-g**.

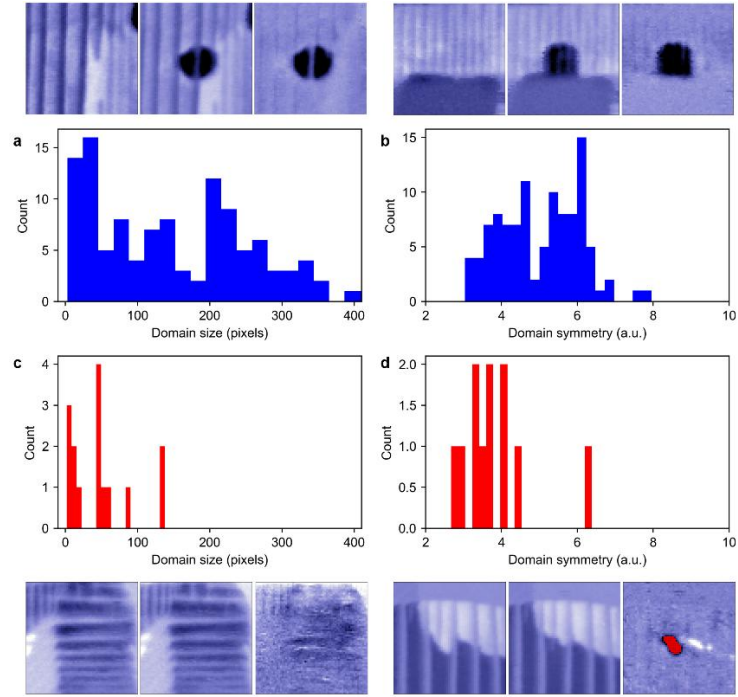

**Figure S2 | Manual comparison between two out-of-plane domains. a-b,** the histograms of **a** poled domain size and **b** poled domain symmetry for all the measured positive out-of-plane domains. The insets above show two characteristic switching processes of positive domains. **c-d,** histograms of **c** poled domain size and **d** poled domain symmetry for all the measured negative out-of-plane domains, with the insets below showing the switching process of negative domains.

3. Data consistency

Figure S3 and S4 show two additional MOB-DKL experiments similar to Figure 3.

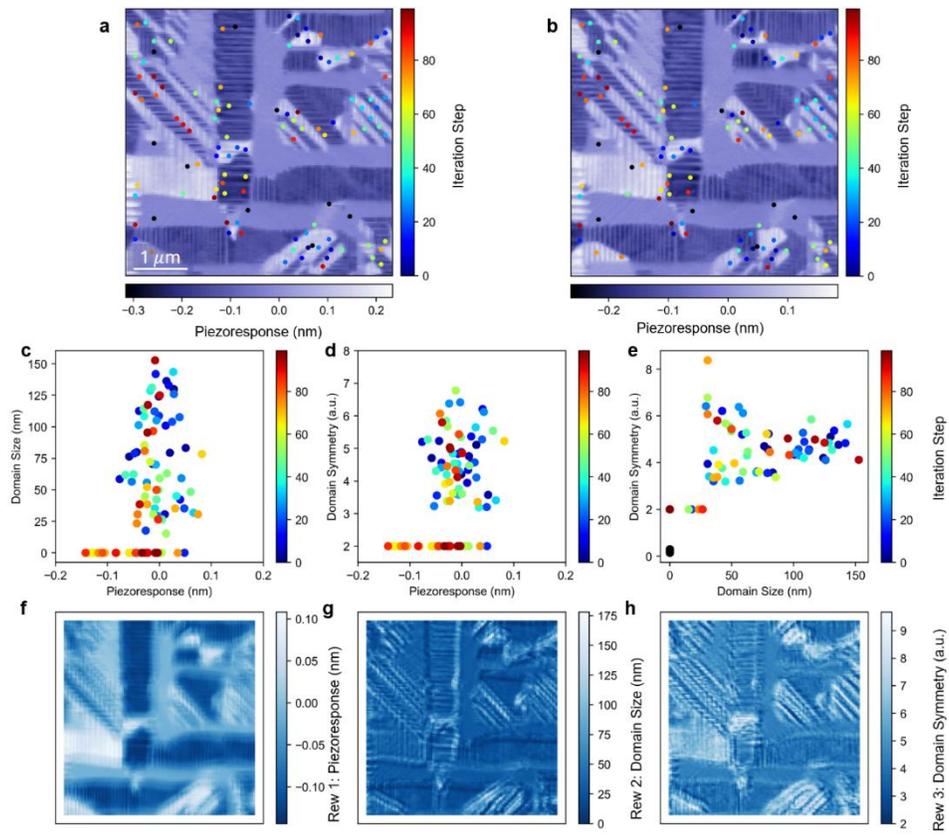

**Figure S3 | More MOBO-DKL experiments on PTKTO.**

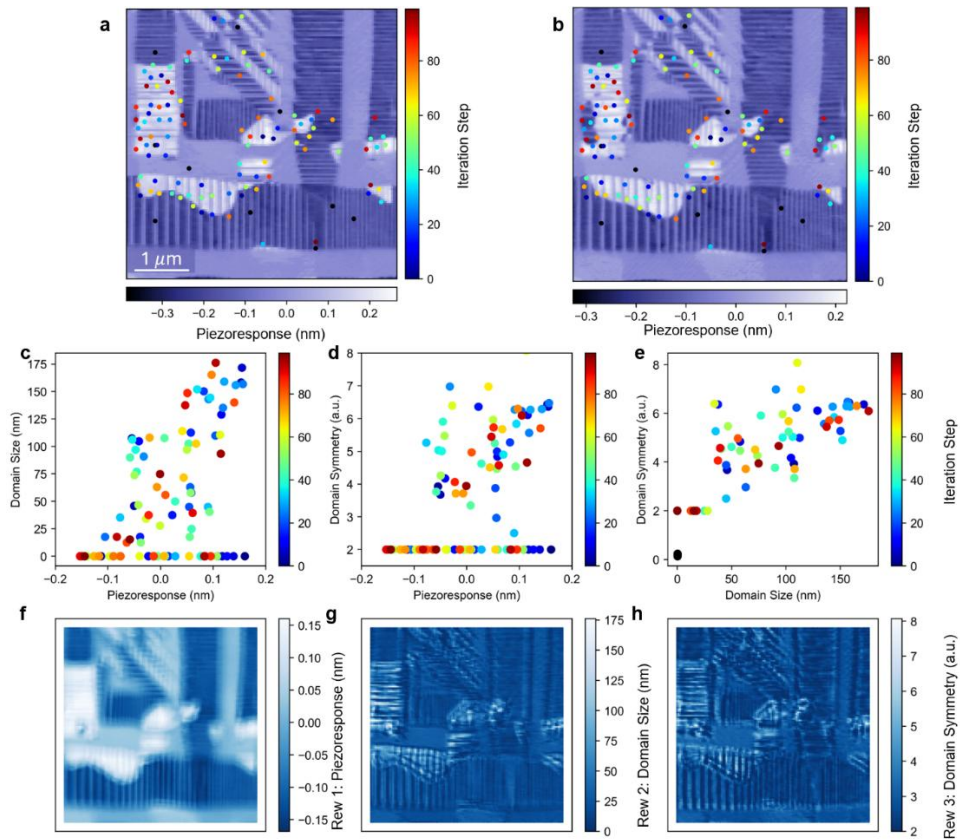

**Figure S4 | More MOBO-DKL experiments on PTKTO.**

**References**

(1) Zhu, J.; Hersam, M. C. Assembly and Electronic Applications of Colloidal Nanomaterials. *Advanced Materials* **2017**, *29* (4), 1603895. DOI: https://doi.org/10.1002/adma.201603895.
(2) Capek, I. *Nanocomposite structures and dispersions*; Elsevier, 2019.
(3) Nataf, G.; Guennou, M.; Gregg, J.; Meier, D.; Hlinka, J.; Salje, E.; Kreisel, J. Domain-wall engineering and topological defects in ferroelectric and ferroelastic materials. *Nature Reviews Physics* **2020**, *2* (11), 634-648.
(4) Chen, S.; Yuan, S.; Hou, Z.; Tang, Y.; Zhang, J.; Wang, T.; Li, K.; Zhao, W.; Liu, X.; Chen, L.; et al. Recent Progress on Topological Structures in Ferroic Thin Films and Heterostructures. *Advanced Materials* **2021**, *33* (6), 2000857. DOI: https://doi.org/10.1002/adma.202000857.
(5) Muralt, P.; Polcawich, R. G.; Trolier-McKinstry, S. Piezoelectric Thin Films for Sensors, Actuators, and Energy Harvesting. *Mrs Bulletin* **2009**, *34* (9), 658-664, Article. DOI: DOI 10.1557/mrs2009.177.

(6) Shaw, T. M.; Trolier-McKinstry, S.; McIntyre, P. C. The properties of ferroelectric films at small dimensions. *Annual Review of Materials Science* **2000**, *30*, 263-298, Review. DOI: DOI 10.1146/annurev.matsci.30.1.263.

(7) Damjanovic, D. Ferroelectric, dielectric and piezoelectric properties of ferroelectric thin films and ceramics. *Reports on Progress in Physics* **1998**, *61* (9), 1267-1324, Review. DOI: 10.1088/0034-4885/61/9/002.

(8) Setter, N.; Damjanovic, D.; Eng, L.; Fox, G.; Gevorgian, S.; Hong, S.; Kingon, A.; Kohlstedt, H.; Park, N. Y.; Stephenson, G. B.; et al. Ferroelectric thin films: Review of materials, properties, and applications. *Journal of Applied Physics* **2006**, *100* (5), Review. DOI: 10.1063/1.2336999.

(9) O'Neill, M. R.; Sessions, D.; Arora, N.; Chen, V. W.; Juhl, A.; Huff, G. H.; Rudykh, S.; Shepherd, R. F.; Buskohl, P. R. Dielectric Elastomer Architectures with Strain-Tunable Permittivity. *Advanced Materials Technologies* **2022**, *7* (11), 2200296. DOI: https://doi.org/10.1002/admt.202200296.

(10) Ohki, Y. Broadband Complex Permittivity and Electric Modulus Spectra for Dielectric Materials Research. *IEEJ Transactions on Electrical and Electronic Engineering* **2022**, *17* (7), 958-972. DOI: https://doi.org/10.1002/tee.23565.

(11) Goshkoderia, A.; Arora, N.; Slesarenko, V.; Li, J.; Chen, V.; Juhl, A.; Buskohl, P.; Rudykh, S. Tunable permittivity in dielectric elastomer composites under finite strains: Periodicity, randomness, and instabilities. *International Journal of Mechanical Sciences* **2020**, *186*, 105880. DOI: https://doi.org/10.1016/j.ijmecsci.2020.105880.

(12) Liu, Y.; Kalinin, S. V. The Power of the Pareto Front: Balancing Uncertain Rewards for Adaptive Experimentation in scanning probe microscopy. 2025; p arXiv:2504.06525.

(13) MacLeod, B. P.; Parlane, F. G. L.; Rupnow, C. C.; Dettelbach, K. E.; Elliott, M. S.; Morrissey, T. D.; Haley, T. H.; Proskurin, O.; Rooney, M. B.; Taherimakhsousi, N.; et al. A self-driving laboratory advances the Pareto front for material properties. *Nature Communications* **2022**, *13* (1), 995. DOI: 10.1038/s41467-022-28580-6.

(14) Low, A.; Lim, Y.-F.; Hippalgaonkar, K.; Vissol-Gaudin, E. Mapping pareto fronts for efficient multi-objective materials discovery. *Authorea Preprints* **2022**.

(15) Mannodi-Kanakkithodi, A.; Pilania, G.; Ramprasad, R.; Lookman, T.; Gubernatis, J. E. Multi-objective optimization techniques to design the Pareto front of organic dielectric polymers. *Computational Materials Science* **2016**, *125*, 92-99.

(16) Rodriguez, B. J.; Nemanich, R. J.; Kingon, A.; Gruverman, A.; Kalinin, S. V.; Terabe, K.; Liu, X. Y.; Kitamura, K. Domain growth kinetics in lithium niobate single crystals studied by piezoresponse force microscopy. *Applied Physics Letters* **2005**, *86* (1). DOI: 10.1063/1.1845594.

(17) He, D. Y.; Qiao, L. J.; Volinsky, A. A.; Bai, Y.; Guo, L. Q. Electric field and surface charge effects on ferroelectric domain dynamics in BaTiO3 single crystal. *Physical Review B* **2011**, *84* (2), 024101, Article. DOI: 10.1103/PhysRevB.84.024101.

(18) Brugère, A.; Gidon, S.; Gautier, B. Abnormal switching of ferroelectric domains created by the tip of an atomic force microscope in a congruent LiTaO3 single-crystal thin film. *Journal of Applied Physics* **2011**, *110* (2), 024102, Article. DOI: 10.1063/1.3607302.

(19) Barakati, K.; Liu, Y.; Funakubo, H.; Kalinin, S. V. Exploring Domain Wall Pinning in Ferroelectrics via Automated High Throughput AFM. 2025; p arXiv:2505.24062.
(20) Roccapriore, K. M.; Dyck, O.; Oxley, M. P.; Ziatdinov, M.; Kalinin, S. V. Automated Experiment in 4D-STEM: Exploring Emergent Physics and Structural Behaviors. *ACS Nano* **2022**, *16* (5), 7605-7614. DOI: 10.1021/acsnano.1c11118
(21) Roccapriore, K. M.; Kalinin, S. V.; Ziatdinov, M. Physics Discovery in Nanoplasmonic Systems via Autonomous Experiments in Scanning Transmission Electron Microscopy. *Adv Sci (Weinh)* **2022**, *9* (36), e2203422. DOI: 10.1002/advs.202203422
(22) Liu, Y.; Kelley, K. P.; Vasudevan, R. K.; Zhu, W.; Hayden, J.; Maria, J. P.; Funakubo, H.; Ziatdinov, M. A.; Trolier-McKinstry, S.; Kalinin, S. V. Automated Experiments of Local Non-Linear Behavior in Ferroelectric Materials. *Small* **2022**, *18* (48), e2204130. DOI: 10.1002/smll.202204130
(23) Liu, Y.; Yang, J.; Vasudevan, R. K.; Kelley, K. P.; Ziatdinov, M.; Kalinin, S. V.; Ahmadi, M. Exploring the Relationship of Microstructure and Conductivity in Metal Halide Perovskites via Active Learning-Driven Automated Scanning Probe Microscopy. *J Phys Chem Lett* **2023**, *14* (13), 3352-3359. DOI: 10.1021/acs.jpclett.3c00223
(24) Liu, Y.; Ziatdinov, M. A.; Vasudevan, R. K.; Kalinin, S. V. Explainability and human intervention in autonomous scanning probe microscopy. *Patterns (N Y)* **2023**, *4* (11), 100858. DOI: 10.1016/j.patter.2023.100858
(25) Kalinin, S. V.; Vasudevan, R.; Liu, Y.; Ghosh, A.; Roccapriore, K.; Ziatdinov, M. Probe microscopy is all you need. *Mach Learn-Sci Techn* **2023**, *4* (2), 023001.
(26) Jesse, S.; Kalinin, S. V.; Proksch, R.; Baddorf, A. P.; Rodriguez, B. J. The band excitation method in scanning probe microscopy for rapid mapping of energy dissipation on the nanoscale. *Nanotechnology* **2007**, *18* (43), 435503. DOI: 10.1088/0957-4484/18/43/435503.

(27) Rodriguez, B. J.; Callahan, C.; Kalinin, S. V.; Proksch, R. Dual-frequency resonance-tracking atomic force microscopy. *Nanotechnology* **2007**, *18* (47). DOI: 10.1088/0957-4484/18/47/475504.

(28) Jesse, S.; Vasudevan, R. K.; Collins, L.; Strelcov, E.; Okatan, M. B.; Belianinov, A.; Baddorf, A. P.; Proksch, R.; Kalinin, S. V. Band excitation in scanning probe microscopy: recognition and functional imaging. *Annu Rev Phys Chem* **2014**, *65*, 519-536, Review; Book Chapter. DOI: 10.1146/annurev-physchem-040513-103609  From NLM Medline.
(29) Kumar, A.; Ehara, Y.; Wada, A.; Funakubo, H.; Griggio, F.; Trolier-McKinstry, S.; Jesse, S.; Kalinin, S. V. Dynamic piezoresponse force microscopy: Spatially resolved probing of polarization dynamics in time and voltage domains. *Journal of Applied Physics* **2012**, *112* (5), 052021, Article. DOI: 10.1063/1.4746080.

(30) Alexe, M.; Gruverman, A.; Harnagea, C.; Zakharov, N. D.; Pignolet, A.; Hesse, D.; Scott, J. F. Switching properties of self-assembled ferroelectric memory cells. *Applied Physics Letters* **1999**, *75* (8), 1158-1160, Article. DOI: 10.1063/1.124628.
(31) Hall, D. A. Review nonlinearity in piezoelectric ceramics. *Journal of Materials Science* **2001**, *36* (19), 4575-4601, Review. DOI: 10.1023/a:1017959111402.

(32) Robert, G.; Damjanovic, D.; Setter, N.; Turik, A. V. Preisach modeling of piezoelectric nonlinearity in ferroelectric ceramics. *Journal of Applied Physics* **2001**, *89* (9), 5067-5074, Article. DOI: Doi 10.1063/1.1359166.
(33) Griggio, F.; Jesse, S.; Kumar, A.; Marincel, D. M.; Tinberg, D. S.; Kalinin, S. V.; Trolier-McKinstry, S. Mapping piezoelectric nonlinearity in the Rayleigh regime using band excitation piezoresponse force microscopy. *Applied Physics Letters* **2011**, *98* (21), 212901. DOI: 10.1063/1.3593138.
(34) Bintachitt, P.; Jesse, S.; Damjanovic, D.; Han, Y.; Reaney, I. M.; Trolier-McKinstry, S.; Kalinin, S. V. Collective dynamics underpins Rayleigh behavior in disordered polycrystalline ferroelectrics. *Proc Natl Acad Sci U S A* **2010**, *107* (16), 7219-7224. DOI: 10.1073/pnas.0913172107
(35) Ovchinnikov, O.; Jesse, S.; Guo, S.; Seal, K.; Bintachitt, P.; Fujii, I.; Trolier-McKinstry, S.; Kalinin, S. V. Local measurements of Preisach density in polycrystalline ferroelectric capacitors using piezoresponse force spectroscopy. *Applied Physics Letters* **2010**, *96* (11), 112906. DOI: 10.1063/1.3360220.
(36) Xu, R.; Liu, S.; Grinberg, I.; Karthik, J.; Damodaran, A. R.; Rappe, A. M.; Martin, L. W. Ferroelectric polarization reversal via successive ferroelastic transitions. *Nature materials* **2015**, *14* (1), 79-86.
(37) Xu, R.; Karthik, J.; Damodaran, A. R.; Martin, L. W. Stationary domain wall contribution to enhanced ferroelectric susceptibility. *Nature communications* **2014**, *5* (1), 3120.
(38) Lin, C.-C.; Hu, Y.; Kim, J.; Lou, D.; Bhat, A.; Kavle, P.; Kim, T. Y.; Dames, C.; Liu, S.; Martin, L. W. Domain-Wall Enhanced Pyroelectricity. *Physical Review X* **2025**, *15* (1), 011063. DOI: 10.1103/PhysRevX.15.011063.
(39) Pratiush, U.; Roccapriore, K. M.; Liu, Y.; Duscher, G.; Ziatdinov, M.; Kalinin, S. V. Building Workflows for Interactive Human in the Loop Automated Experiment (hAE) in STEM-EELS. 2024; p arXiv 2404.07381.
(40) Liu, Y.; Ziatdinov, M. A.; Vasudevan, R. K.; Kalinin, S. V. Explainability and human intervention in autonomous scanning probe microscopy. *Patterns* **2023**, *4* (11), 100858. DOI: https://doi.org/10.1016/j.patter.2023.100858.
(41) Liu, Y.; Kelley, K. P.; Vasudevan, R. K.; Zhu, W.; Hayden, J.; Maria, J.-P.; Funakubo, H.; Ziatdinov, M. A.; Trolier-McKinstry, S.; Kalinin, S. V. Automated Experiments of Local Non-Linear Behavior in Ferroelectric Materials. *Small* **2022**, *18* (48), 2204130. DOI: https://doi.org/10.1002/smll.202204130.
(42) Gardner, J.; Pleiss, G.; Weinberger, K. Q.; Bindel, D.; Wilson, A. G. Gpytorch: Blackbox matrix-matrix gaussian process inference with gpu acceleration. *Advances in neural information processing systems* **2018**, *31*.
(43) Balandat, M.; Karrer, B.; Jiang, D. R.; Daulton, S.; Letham, B.; Wilson, A. G.; Bakshy, E. BoTorch: A Framework for Efficient Monte-Carlo Bayesian Optimization. 2019; p arXiv 1910.06403.
(44) Liu, Y.; Pratiush, U.; Bemis, J.; Proksch, R.; Emery, R.; Rack, P. D.; Liu, Y.-C.; Yang, J.-C.; Udovenko, S.; Trolier-McKinstry, S.; et al. Integration of scanning probe microscope with high-performance computing: Fixed-policy and reward-driven workflows implementation. *Review of Scientific Instruments* **2024**, *95* (9). DOI: 10.1063/5.0219990

(45) Williams, C. K.; Rasmussen, C. E. *Gaussian processes for machine learning*; MIT press Cambridge, MA, 2006.

(46) Wilson, A. G.; Hu, Z.; Salakhutdinov, R.; Xing, E. P. Deep Kernel Learning. In Proceedings of the 19th International Conference on Artificial Intelligence and Statistics, Proceedings of Machine Learning Research; 2016.

(47) Paria, B. Strategies for Black-Box and Multi-Objective Optimization. PhD thesis, Carnegie Mellon University, 2022.

(48) Daulton, S.; Balandat, M.; Bakshy, E. Parallel Bayesian Optimization of Multiple Noisy Objectives with Expected Hypervolume Improvement. 2021; p arXiv:2105.08195.

(49) Daulton, S.; Balandat, M.; Bakshy, E. Differentiable Expected Hypervolume Improvement for Parallel Multi-Objective Bayesian Optimization. 2020; p arXiv:2006.05078.